\documentstyle[12pt,epsf]{article}

 \large\normalsize

\newcommand{\<}{\langle}
\renewcommand{\>}{\rangle}
\newcommand{\be}{\begin{equation}}
\newcommand{\ee}{\end{equation}}
\newcommand{\tr}{\mathop{\rm Tr}\nolimits}
\def\spose#1{\hbox to 0pt{#1\hss}}
\def\ltapprox{\mathrel{\spose{\lower 3pt\hbox{$\mathchar"218$}}
 \raise 2.0pt\hbox{$\mathchar"13C$}}}
\def\gtapprox{\mathrel{\spose{\lower 3pt\hbox{$\mathchar"218$}}
 \raise 2.0pt\hbox{$\mathchar"13E$}}}
\def\reff#1{(\ref{#1})}
\def\bgamma{\vec{\mbox{$\gamma$}}}

\def\bsig{\vec{\mbox{$\sigma$}}}

\def\bx{x}
\def\by{y}
\def\bun{e}
\newcommand{\1}{1\!\!\!\bot}
\newcommand{\ba}{\begin{eqnarray}}
\newcommand{\ea}{\end{eqnarray}}

\begin{document}

\title{Numerical Study of the \\ Fundamental Modular Region \\
   in the Minimal Landau Gauge}

\author{{\small Attilio Cucchieri}\thanks{~E-mail address:
{\tt cucchieri@roma2.infn.it}.} \\ [-0.1cm]
  {\small Gruppo APE -- Dipartimento di Fisica} \\ [-0.1cm]
  {\small Universit\`a di Roma ``Tor Vergata'' } \\ [-0.1cm]
  {\small Via della Ricerca Scientifica 1, I-00133 Roma, ITALY}}

\date{\today}

\maketitle

\begin{abstract}
We study numerically the so-called fundamental modular region
$\Lambda$, a region free of Gribov copies, in the
minimal Landau gauge for pure $SU(2)$ lattice gauge theory.
To this end we evaluate the influence of Gribov copies
on several quantities --- such as the smallest
eigenvalue of the Faddeev-Popov matrix,
the third and the fourth derivatives
of the minimizing function, and
the so-called horizon function ---
which are used to characterize the region $\Lambda$.
Simulations are done at four different values of the coupling:
$\beta = 0\mbox{,}\, 0.8\mbox{,}\,
1.6\mbox{,}\, 2.7\,$, and for volumes up to $16^4$.
We find that typical (thermalized and gauge-fixed)
configurations, including those belonging to the region $\Lambda$,
lie very close to the Gribov horizon $\partial \Omega$,
and are characterized, in the limit of large lattice volume,
by a negative-definite horizon tensor.
\end{abstract}



\section{Introduction}

Gauge theories, being invariant under local gauge transformations,
are systems with redundant dynamical variables, which
do not represent true dynamical degrees of freedom.
The objects of interest are not the gauge fields
themselves, but rather the classes (orbits) of gauge-related fields.
The elimination of such redundant gauge degrees of freedom is essential
for understanding and extracting physical information from these
theories. This is usually done by introducing
a gauge-fixing condition
which determines a representative gauge field on each orbit.
In reference \cite{Gr} Gribov showed that Coulomb and Landau gauge-fixing
conditions do not fix the gauge fields uniquely, i.e.\
there are many gauge-equivalent configurations
satisfying the Coulomb or Landau
transversality condition. These {\em Gribov
copies} do not affect perturbative calculations, but
their elimination could play a crucial role for
non-perturbative features of gauge theories.
Gribov's result was generalized by Singer \cite{Singer} to the
case of a generic compact semi-simple non-abelian Lie group.
In this work the author considered continuous gauge fixing, and
boundary conditions (i.e.\ value of the gauge
fields at infinity) such that the
euclidean space-time is compactified to the
four-dimensional sphere $S^{4}$. A similar
analysis was done by Killingback \cite{Killing}
for the case of periodic
boundary conditions (i.e.\ the four-dimensional
torus $T^{4}$), which are the
boundary conditions usually employed in lattice gauge theory.

A possible solution to the problem of Gribov copies
is to restrict the functional integral
to a subset of the gauge-field space, the {\em fundamental modular
region} $\Lambda$, which is the set of absolute minima
of a Morse function on the gauge orbits \cite{Morse,vanbaal1}.
In the so-called {\em minimal Landau gauge} this
Morse function is defined as
\be
{\cal E}_{A}[ g ]\,\equiv\,\|\,A^{(g)}\,\|^{2}\,
\equiv\,\frac{1}{2}\,\sum_{\mu\mbox{,}\,a}\,
\int\,d^{d}x\,\left\{\,\left[\,A_{\mu}^{(g)}\,\right]^{a}(x)\,\right\}^{2}
\;\mbox{.}
\label{eq:Econt}
\ee
It has been proven \cite{vanbaal1,STSF,DAZ,PvB} that,
with this choice of the Morse function,
every orbit intersects the interior of the fundamental
modular region once and only once, i.e.\
in the {\em interior} of $\Lambda$ the absolute minima are
non-degenerate and there are {\em no} Gribov copies.
However, on the {\em boundary} of the fundamental modular region
there are degenerate absolute minima,
and only when they have been identified can we obtain a
region truly free of Gribov copies \cite{vanbaal1,PvB}.

\vspace*{0.2cm}

The problem of Gribov copies is also present in the
lattice regularization of gauge theories \cite{MPR,PFH}.
Although this formulation does not
require gauge fixing, due to asymptotic freedom, the continuum
limit is the weak-coupling limit, and a weak-coupling expansion
requires gauge fixing. Thus, gauge-dependent quantities are
usually introduced on the lattice, and Gribov copies have to be taken into
account in lattice gauge theory as well.

A fundamental modular region can be defined also on the
lattice. In this case, for the minimal Landau gauge, we
consider the absolute minima of the functional\footnote{~This
definition applies to $SU(2)$ lattice gauge theory
in $4$ dimensions.  We consider a standard Wilson action
with periodic boundary conditions, and lattice volume
$V$. For notations we refer to \cite{Gnoise}.}
\be
{\cal E}_{U}[ g ] \; \equiv \;
        \frac{1}{8\,V} \sum_{\bx} \sum_{\mu} \,
            \tr \, \left[ \, \1 \, -
\, g(\bx) \;
          U_{\mu}(\bx) \;
                   g^{\dagger}(\bx + \bun_{\mu})
                \, \right]
\label{eq:Ewith1}
\;\mbox{,}
\ee
which is the lattice analogue of the Morse function
${\cal E}_{A}[ g ]$ used in the continuum
[see eq.\ \reff{eq:Econt}]. Since the
gauge orbit is a compact manifold on a finite
lattice, this functional is bounded
and the existence of an absolute minimum
for ${\cal E}_{U}$ follows immediately.
Let us notice that the functional ${\cal E}_{U}[ g ]$ can
be rewritten \cite{Grab} as the
quadratic form\footnote{~In equations
\reff{eq:EwithD} and \reff{eq:EwithD2} we use $g$
to indicate four-dimensional unit vectors.}
\be
{\cal E}_{U}[ g ] \, = \,
   \frac{1}{8\,V} \sum_{\bx} \,
g(\bx)\,\cdot\, \left(\,\nabla_{U}\,g\,\right)(\bx)
\;\mbox{,}
\label{eq:EwithD}
\ee
where
\be
\left(\,\nabla_{U}\,g\,\right)(\bx)
\;=\,
\sum_{\mu}\,
\left[ \,2\,g^{\dagger}(\bx)\, -\,
U_{\mu}(\bx) \;
                   g^{\dagger}(\bx + \bun_{\mu})
\;-\;
U_{\mu}^{\dagger}(\bx - a \bun_{\mu}) \;
                   g^{\dagger}(\bx - \bun_{\mu})
\,\right]
\label{eq:EwithD2}
\ee
is the gauge-covariant Laplacian. The minimization of
a quadratic form is a standard and simple problem
if the variables are elements of
a linear space \cite{PTVF}. In our case, however,
these variables
are $SU(2)$ matrices (i.e.\ 4-dimensional unit vectors),
and due to this non-linearity the numerical
search for the absolute minimum becomes highly non-trivial.

Most of the properties proved in the 
continuum for the fundamental modular region 
can be extended to the lattice case
\cite{Z1,Z2}.
In particular, an explicit example of degenerate
absolute minima on the boundary of $\Lambda$ is given in
reference \cite{Z1}. 

\vspace*{0.2cm}

In this work we want to {\em characterize}
the fundamental modular region
by evaluating numerically
diagnostic quantities (see Section \ref{Sec:FMRstudio})
at relative and absolute minima.
To date, relatively few studies
\cite{PFH,Gnoise,FH} have been done in order
to analyze the influence of Gribov copies ({\em Gribov noise})
on some lattice quantities. However, these numerical studies
have never tried to characterize the ``geometry'' of
the orbit space. We know 
general properties of the fundamental
modular region, and we know particular cases
in which this region can be studied
analytically \cite{vanbaal1,STSF,DAZ,PvB,Z1,Z2,vanbaal2}.
On the contrary, we do not know
what happens in numerical simulations. This work
aims at filling this gap, and at providing information
about the ``localization'' in the gauge-field space 
of a {\em typical} thermalized gauge-fixed configuration.
Preliminary results have been reported in \cite{Athesis}.


\section{Lattice Landau Gauge-Fixing Condition}
\label{Sec:lattice}

In this section we analyze in more detail
the lattice Landau gauge-fixing condition. Let us recall
that this gauge condition is imposed
by minimizing the functional ${\cal E}_{U}$,
defined in eq.\ \reff{eq:Ewith1}, with respect to
the variables $\{ g(\bx) \}$, keeping the
thermalized configuration $\{ U_{\mu}(\bx) \}$ fixed.

We consider a one-parameter subgroup
$g(\tau; \bx )$ of $SU(2)$ defined by
\be
g(\tau; \bx) \equiv
   \exp \left[ \;i\, \tau \, \bgamma(\bx) \cdot \bsig \; \right]
\label{eq:oneparam}
\;\mbox{;}
\ee
here the parameter $\tau$ is a real number,
$\bgamma(\bx)$ is a three-dimensional real vector, and
the components of $\bsig$ are the three Pauli matrices.
Then the functional ${\cal E}_{U}$, defined
in eq.\ \reff{eq:Ewith1}, can be regarded as a function of the
parameter $\tau$, and its first derivative
--- with respect to $\tau$ and at $\tau = 0$ ---
is given by the well-known expression
\be
{\cal E}^{'}(0) \, = \, 
  \frac{1}{4\,V} \, \sum_{\bx} \, \sum_{a} \,
      \gamma^{a}(\bx)  \, 
          \left(\nabla\cdot
        A^{\left( g \right)}\right)^{a}(\bx)
  \; \mbox{,}
\label{eq:Ederiv}
\ee 
where
\be
\left(\nabla\cdot A \right)^{a}(\bx) \equiv
  \sum_{\mu} \, A_{\mu}^{a} (\bx) -
                  A_{\mu}^{a} (\bx - \bun_{\mu})
\label{eq:diverA}
\ee
is the lattice divergence of the gluon field
$\, A_{\mu}^{a}(\bx) \,$.
If $\{ U_{\mu}\left(\bx\right) \}$ is a stationary point of
${\cal E}(\tau)$ at $\tau = 0$ 
then we have
${\cal E}^{'}(0) = 0$ for every $\{ \gamma^{a} (\bx) \}$. This
implies
\be
\left(\nabla \cdot A \right)^{a}(\bx) = 0 
\label{eq:diverg0}
\ee
for any $\bx$ and $a$,
which is the
lattice formulation of the usual Landau gauge-fixing condition
in the continuum.

\begin{figure}
\epsfxsize=0.9\textwidth
\vskip 2cm
\centerline{\epsffile{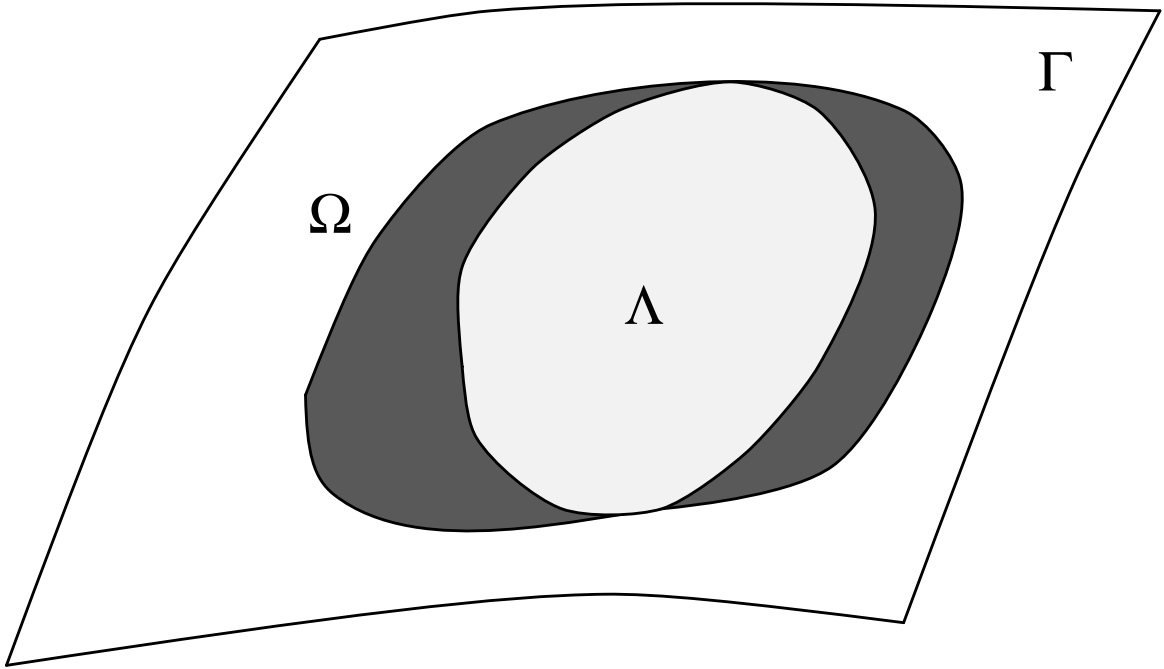}}
\vskip 1cm
\caption{~The plane of the figure represents the hyperplane
         $\Gamma$ of transverse configurations ($\nabla \cdot A = 0$).
         It contains the Gribov region $\Omega$ defined in eq.\ 
         \protect\reff{eq:Omega}, and the fundamental modular
         region $\Lambda$. Note that parts of the boundaries
         $\partial \Omega$ and $\partial \Lambda$ are in common.}
\label{fig:regioni}
\end{figure}

Still considering the one-parameter subgroup defined in eq.\
\reff{eq:oneparam}, we can evaluate
the second derivative of ${\cal E}(\tau)$.
Following references \cite{Z1,Z2} one can check that
\ba
{\cal E}^{''}(0)&=&\frac{1}{4\,V}\sum_{\bx\mbox{,}\,\by}\,
   \sum_{a\mbox{,}\,b}\,\gamma^{a}(\bx)\,
{\cal M}_{U}^{a\, b}(\bx\mbox{,}\by)\,
\gamma^{b}(\by) \label{eq:secderofE2}\\
& \equiv & \,\frac{1}{4\,V}\sum_{\bx}\,
   \sum_{\mu\mbox{,}\, a}\,\biggl\{
          \Bigl[\,\gamma^{a}(\bx + \bun_{\mu})\,-\,
                       \gamma^{a}(\bx)\,\Bigr]^{2}\,U_{\mu}^{0}(\bx)
\nonumber \\
          & & \,+\,
  \sum_{b\mbox{,}\,c}\,\epsilon^{a b c}\,
   \Bigl[\,\gamma^{a}(\bx + \bun_{\mu})\,-\,
                       \gamma^{a}(\bx)\,\Bigr]  \,
   \Bigl[\,\gamma^{b}(\bx + \bun_{\mu})\,+\,
                       \gamma^{b}(\bx)\,\Bigr] \,A^{c}_{\mu}(\bx)
\biggl\}
\;\mbox{,}
\label{eq:secderofE}
\ea
where ${\cal M}_{U}^{a\, b}(\bx\mbox{,}\by)$
is the lattice Faddeev-Popov matrix
and $\epsilon^{a b c}$ is the complete anti-symmetric tensor.
It is clear that this second derivative is
null if the vector $\gamma^{a}(\bx)$
is constant, i.e.\ the
Faddeev-Popov matrix has a
trivial null eigenvalue corresponding to a constant eigenvector.
From eq.\ \reff{eq:secderofE2} we obtain that,
if $\{ U_{\mu}\left(\bx\right) \}$ is a local minimum of
${\cal E}(\tau)$ at $\tau = 0$, then the matrix ${\cal M}_{U}$ is
positive definite (in the subspace orthogonal to the space of
constant vectors). This implies that any local minimum
of the functional ${\cal E}_{U}$ belongs to the region
(see Figure \ref{fig:regioni})
\be
\Omega \equiv \left\{\, A \, : \, \nabla \cdot A \,=\,0\mbox{,}\,\,
    {\cal M}_{U}\,\geq\,0\, \right\}
\label{eq:Omega}
\;\mbox{.}
\ee
This region was introduced by Gribov \cite{Gr}
in the attempt of getting rid
of spurious gauge copies. It is delimited
by the so-called {\em first Gribov horizon} $\partial \Omega$, i.e.\ the set of
configurations for which the smallest nontrivial
eigenvalue $\lambda_{min}$ of the Faddeev-Popov matrix is zero.
Clearly the Gribov region $\Omega$ includes the fundamental modular
region $\Lambda$. However, there are points ---
the so-called {\em singular boundary} points \cite{vanbaal1,
PvB,Z1,Z2,vanbaal2} --- on the boundary
of $\Lambda$ which are also on the boundary of $\Omega$ (see Figure \ref{fig:regioni}).
A typical example are degenerate absolute minima of
the Morse function for which the degeneracy is continuous.
In fact, as said in the Introduction, {\em all} the degenerate absolute
minima are found on the boundary of the fundamental
modular region. In the case of a
continuous degeneracy, the Faddeev-Popov operator {\em must}
have a zero eigenvalue, i.e.\ these minima are {\em also} on the
boundary of $\Omega$.
An explicit example of singular boundary points on the
lattice has been found by Zwanziger (see
\cite[Appendix E]{Z1}).

Finally, we can evaluate the third and fourth derivatives
of ${\cal E}(\tau)$ with respect to $\tau$, at $\tau = 0$. Following again
\cite{Z1,Z2} we obtain\footnote{~Our notation
is slightly different from the notation
used in references \cite{Z1} and \cite{Z2}. This explains the
difference between the coefficients in formulae \reff{eq:E3} and
\reff{eq:E4} and the coefficients in the corresponding equations
in references \cite{Z1,Z2}.}
\ba
{\cal E}^{'''}(0)&=&\frac{3}{4\,V}\sum_{\bx}\,\biggl\{
\left[\,{\vec{\mbox{$\gamma$}}}^{2}(\bx + \bun_{\mu})\,-\,
                       {\vec{\mbox{$\gamma$}}}^{2}(\bx)\,\right]
\,\sum_{\mu\mbox{,}\, a}\, \Bigl[\,\gamma^{a}(\bx + \bun_{\mu})\,+\,
                       \gamma^{a}(\bx)\,\Bigr] \,A^{a}_{\mu}(\bx)\,
\biggl\} \nonumber \\
  & & \qquad\,\,\,\,\,\,\,-\,\frac{1}{V}\sum_{\bx}\,
     \biggl\{\,{\vec{\mbox{$\gamma$}}}^{2}(\bx)\,\sum_{a}\,\Bigl[\,
      \gamma^{a}(\bx)\,\left(\nabla\cdot A\right)^{a}(\bx)\,\Bigr]\,
\biggl\}
\label{eq:E3}
\ea
and
\ba
{\cal E}^{''''}(0)&=&\frac{3}{4\,V}\sum_{\bx}\,\sum_{\mu}\,
    \biggl\{\Bigl[\,{\vec{\mbox{$\gamma$}}}^{2}(\bx + \bun_{\mu})
\,-\, {\vec{\mbox{$\gamma$}}}^{2}(\bx)\,\Bigr]^{2}\,
U_{\mu}^{0}(\bx)\,\Biggl\} \nonumber \\
& & \quad -\,\frac{1}{V}\sum_{\bx\mbox{,}\, \by}\,
     \biggl\{\,{\vec{\mbox{$\gamma$}}}^{2}(\bx)
\,\sum_{a \mbox{,}\, b}
   \,\Bigl[\,\gamma^{a}(\bx)\,
{\cal M}_{U}^{a\, b}(\bx\mbox{,}\,\by)\, \gamma^{b}(\by)\,\Bigr]
    \,\biggl\}
\;\mbox{.}
\label{eq:E4}
\ea


\section{Characterization of the Fundamental Modular Region}
\label{Sec:FMRstudio}

In order to characterize the ``localization'' of the gauge-fixed
configuration in the hyperplane $\Gamma$ of transverse
configurations ($\nabla \cdot A = 0$; see Figure \ref{fig:regioni}),
we consider the smallest nonzero eigenvalue $\lambda_{min}$
of the Faddeev-Popov matrix ${\cal M}_{U}$, and its
corresponding eigenvector $\omega^{a}_{min}(\bx)$,
i.e.\ we solve the eigenvalue problem
\be
\sum_{\by}\,\sum_{b}\,
{\cal M}_{U}^{a\, b}(\bx\mbox{,}\,\by)\, \omega^{b}_{min}(\by)
\,=\,\lambda_{min}\,\omega^{a}_{min}(\bx)
\label{eq:lambdamin}
\;\mbox{.}
\ee
As already said, the matrix ${\cal M}_{U}$ has a trivial
null eigenvalue corresponding to a constant eigenvector.
Therefore, this equation has to be solved in the subspace
orthogonal to constant vectors, namely the 
eigenvector $\omega^{a}_{min}(\bx)$ should satisfy
the relation $\sum_{\bx}\,\omega^{a}_{min}(\bx)\,=\,0$.
As for the eigenvalue $\lambda_{min}$,
we know that it is positive,
since any local minimum of ${\cal E}_{U}$ belongs to $\Omega$.
Moreover, for the vacuum configuration
$U_{\mu}(\bx) \,=\, \1\,$ (which also belongs
to $\Omega$ and to $\Lambda$), the Faddeev-Popov matrix is simply (minus)
the lattice Laplacian\footnote{~This can be seen from
equation \reff{eq:secderofE}
with $U^{0}_{\mu}(\bx) = 1$ and $A_{\mu}^{c}(\bx) = 0$.}. Therefore,
in this case, the value of $\lambda_{min}$ is given by
\be
\lambda_{Lap}\,=\,
4 \sin^{2}\left(\,\pi / N\,\right)
\label{eq:lLap}
\;\mbox{,}
\ee
where $N$ is the lattice size.
Finally, this eigenvalue goes to zero as the first Gribov
horizon $\partial \Omega$ is approached. So, the value of $\lambda_{min}$ can
be interpreted as a sort of distance between the minimum
and $\partial \Omega$, and we may test whether its value is
largest (in average) for the absolute minimum.

\begin{figure}
\epsfxsize=1.3\textwidth
\vskip -5cm
\centerline{\epsffile{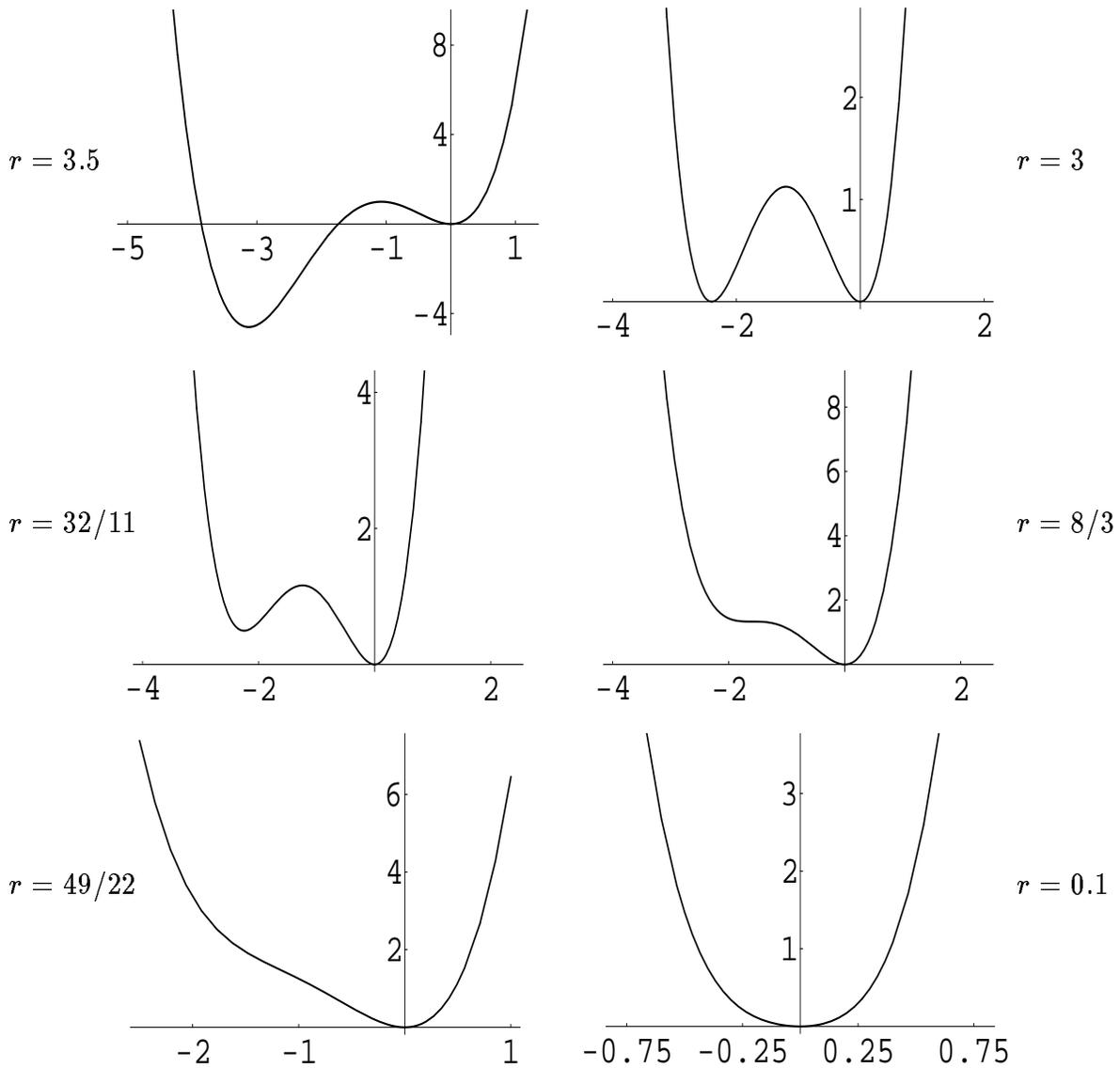}}
\vskip -5cm
\caption{~Plot of $\,{\cal E}(\tau) \,-\, {\cal E}(0)\,$ as
         a function of $\tau$ for six different values of
         the ratio $r$ [see equations
         \protect\reff{eq:ratiodefinizione} and
         \protect\reff{eq:Eofr}].
         For each value of $r$, the values for
         ${\cal E}^{''}(0)$ and ${\cal E}^{'''}(0)$,
         and the scale on the $y$ axis have been chosen ``ad hoc''
         in order to emphasize the differences
         between the various cases. Recall that we are
         sitting at the minimum at $\tau = 0$.}
\label{fig:ratior}
\end{figure}

After we have evaluated the eigenvector 
$\omega^{a}_{min}(\bx)$, we can set $\gamma^{a}(\bx)\,=\,
\omega^{a}_{min}(\bx)$ in the one-parameter subgroup defined
in eq.\ \reff{eq:oneparam}.
Then, if $\{\,U_{\mu}(\bx)\,\}$ is the configuration which minimizes
the functional ${\cal E}_{U}$,
we can study the behavior of the minimizing functional
near the minimum using the gauge
transformation generated by this one-parameter subgroup.
More exactly, we can
analyze its behavior along the ``direction'' of 
$\omega^{a}_{min}(\bx)$, i.e.\ along the direction for
which the rate of increase of the functional ${\cal E}_{U}$
is smallest. To this end we expand
${\cal E}(\tau)$ in powers of $\tau$ around
the minimum, i.e.\ around $\tau = 0$. We then write
\be
{\cal E}(\tau)\,=\,{\cal E}(0)\, +\,\frac{\tau^{2}}{2}\,
\left[\, {\cal E}^{''}(0) \, +\, \frac{\tau}{3}\,
{\cal E}^{'''}(0)\,+\, \frac{\tau^{2}}{12}\,
    {\cal E}^{''''}(0)\,\right]\,+\,\ldots
\label{eq:aroundmin}
\;\mbox{,}
\ee
where the derivatives
of ${\cal E}(\tau)$ [see equations 
\reff{eq:secderofE}, \reff{eq:E3} and \reff{eq:E4}]
are evaluated with $\gamma^{a}(\bx)\,=\,\omega^{a}_{min}(\bx)$,
and the eigenvector $\omega^{a}_{min}(\bx)$ is
normalized to one. We can now define the ratio
\be
r\,\equiv\,\frac{\left[\,{\cal E}^{'''}(0)\,\right]^{2}}{
           {\cal E}^{''}(0) \, {\cal E}^{''''}(0)}
\label{eq:ratiodefinizione}
\;\mbox{,}
\ee
which is independent of the scale $\tau$, and
rewrite equation \reff{eq:aroundmin} as
\be
{\cal E}(\tau)\,=\,{\cal E}(0)\, +\,
\frac{\tau^{2}\,{\cal E}^{''}(0)}{2}\,
\left[\,1\, +\, \frac{1}{3}\,\frac{\tau\,{\cal E}^{'''}(0)}{
{\cal E}^{''}(0)}\,+\, \frac{1}{12\, r}\,
\left(\,\frac{\tau\,{\cal E}^{'''}(0)}{
{\cal E}^{''}(0)}\,\right)^{2} \,\right]
\,+\,\ldots
\label{eq:Eofr}
\;\;\mbox{.}
\ee
If we make the change of variables $x \equiv \tau\,
{\cal E}^{'''}(0) / {\cal E}^{''}(0)$, then it is
clear that the shape of the minimizing function
around its minimum is fixed by the value of
the ratio $r$. As an example, in Figure \ref{fig:ratior},
we show the behavior of $\,{\cal E}(\tau) \,-\,
{\cal E}(0)\,$ for six different values of the ratio
$r$. Let us notice that, for $r > 8/3$, there
are Gribov copies of the minimum at $\tau = 0$;
in particular, there is a maximum, which
does not belong to the Gribov region $\Omega$,
and a second minimum, which is an example of
a Gribov copy inside the first Gribov horizon $\partial \Omega$.
Thus, we expect a value of $r$ smaller (in average)
for the absolute minimum than for a
generic relative minimum.
Let us also notice that
plots in Figure \ref{fig:ratior}
are related to the bifurcation process described
in reference \cite{vanbaal1}. In that
case the author was following a stationary
point of the minimizing function, while moving
from inside to outside the region $\Omega$. 
Here, on the contrary, we sit at the minimum
at $\tau = 0$ and look at its surroundings
for different values of the ratio $r$. We note that
changing the value of $r$ is equivalent to 
moving this minimum inside the region $\Omega$.  

If we set 
$\gamma^{a}(\bx)\,=\,\omega^{a}_{min}(\bx)$,
and $\omega^{a}_{min}(\bx)$ is normalized to $1$, then
from formulae \reff{eq:secderofE2} and \reff{eq:lambdamin} we
obtain $\lambda_{min} \,=\,4\,V\,{\cal E}^{''}(0)$.
In our simulations we evaluate the eigenvalue
$\lambda_{min}$, the third and
fourth derivatives of the minimizing 
function\footnote{~Also these derivatives
are evaluated at $\tau = 0$ and for
$\gamma^{a}(\bx)\,=\,\omega^{a}_{min}(\bx)\,$
[see eq.\ \protect\reff{eq:E3} and \protect\reff{eq:E4}],
and multiplied by $4\,V$.},
and the ratio $r$.

Recently Zwanziger proposed \cite{Z1,Z2} a modification of the
$SU(N)$-Yang-Mills action
which effectively constrains the functional integral to the fundamental
modular region $\Lambda$
in minimal Landau gauge. This effective action
is given by
\be
S_{eff}[ U ] \,=\, \beta\,S_{W}[ U ] + \alpha\,H[ U ]
\label{eq:Seff}
\;\mbox{,}
\ee
where $\beta\,S_{W}[ U ]$ is the standard Wilson action,
$\alpha$ is a new thermodynamic parameter and
$H[ U ]$ is the so-called {\em horizon function} defined as
\be
H[ U ] \,\equiv\,\frac{1}{12}
\,\sum_{\mu\mbox{,}\,a}\,H_{\mu\,a\mbox{,}\,\mu\,a}[ U ]
\label{eq:hfunc}
\;\mbox{.}
\ee
Here $H_{\mu\,a\mbox{,}\,\nu\,b}[ U ]$
is the {\em horizon tensor} given by
\be
H_{\mu\,a\mbox{,}\,\nu\,b}[ U ] \equiv
\sum_{\bx\mbox{,}\,\by}\,
   \sum_{c\mbox{,}\,d}\,\biggl\{\,B^{c}_{\mu\,a}(\bx)
\,\left(\,
{\cal M}_{U}^{- 1}\,\right)^{c\, d}(\bx\mbox{,}\,\by)\,
B^{d}_{\nu\,b}(\by)
\,\biggl\}
\, -\,\delta_{\mu\,\nu}\,\delta_{a\,b}\,
    \sum_{\bx}\,U_{\mu}^{0}(\bx)
\label{eq:htensor}
\ee
and
\be
B^{c}_{\mu\,a}(\bx) \equiv
\delta^{c}_{a}\,\left[\,
U^{0}_{\mu}(\bx) \,-\,U^{0}_{\mu}(\bx - a \bun_{\mu})\,\right] 
\, +\,\epsilon^{\phantom{a} b c}_{a}\left[\,
A^{b}_{\mu}(\bx) \,+\,A^{b}_{\mu}(\bx - a \bun_{\mu})\,\right]
\;\mbox{.}
\label{eq:Bvector}
\ee
With this new action the standard Yang-Mills theory is recovered
(in the infinite-volume limit)
only when the thermodynamic parameter $\alpha$
has the critical value $\alpha_{cr}$ fixed by the
so-called {\em horizon condition}\footnote{~Here
$\< . \>$ indicates an expectation value calculated
in an ensemble $Z(\alpha)$ depending on the thermodynamic
parameter $\alpha$. For a definition of $Z(\alpha)$
see \cite[eq.\ (1.14)]{Z2}.}
\be
- \alpha_{cr}^{- 1} \, = \, \lim_{V \to \infty}
\frac{1}{V} \, \<\, H[ U ]\, \>
\label{eq:horcondit}
\;\mbox{.}
\ee

For the vacuum configuration
$U_{\mu}(\bx) \,=\, \1\,$ we obtain
$B^{c}_{\mu\,a}(\bx) = 0$, and therefore the horizon tensor 
is diagonal
and equal to $-\,\delta_{\mu\,\nu}\,\delta_{a\,b}\,V$.
On the contrary,
at the first Gribov horizon $\partial \Omega$, the term containing
the inverse of the Faddeev-Popov matrix blows up,
i.e.\ both the horizon tensor and the horizon function
are positive and become larger and larger as we
approach $\partial \Omega$.
In particular, points on the boundary of $\Omega$
where the horizon function is infinite
can be explicitly exhibited \cite{Z3}.
However, the horizon function is not necessarily infinite
for all configurations on the boundary of $\Omega$.
In fact we can rewrite this function as \cite{Z2}
\be
H[ U ] \,=\,\frac{1}{12}\,\sum_{\mu\mbox{,}\,a}\,\sum_{n}\,
\frac{1}{\lambda_{n}}\,
\left[\, \sum_{\bx\mbox{,}\,c}\,B^{c}_{\mu\,a}(\bx)\,
\omega_{n}^{c}(\bx)\,\right]^{2}
\, -\,\frac{1}{4}\,
    \sum_{\bx}\,\sum_{\mu}\, U_{\mu}^{0}(\bx)
\;\mbox{,}
\label{eq:hfunct2}
\ee
where $\lambda_{n}$ and $\omega_{n}^{c}(\bx)$ are, respectively,
the eigenvalues and the corresponding
normalized eigenvectors of the 
Faddeev-Popov matrix ${\cal M}_{U}$. Then,
for the so-called {\em degenerate gauge orbits} \cite{Z2},
one obtains that the absolute minimum is degenerate,
namely it is on the common boundaries of $\Lambda$ and
$\Omega$, and that the term of the horizon function which
is proportional to $ 1 / \lambda_{min}$ is of the
indeterminate\footnote{~This result follows from the relation
$ B^{c}_{\mu\,a}(\bx) \,=\, -\,\left( \, D_{\mu}^{\dagger}[ U ]
\,\right)^{c}_{a}$,
where $D_{\mu}[ U ]$ is the lattice gauge-covariant derivative, 
and the fact that degenerate gauge orbits have a nonzero solution
for the equation $D_{\mu}[ U ]\,\omega \, =\, 0$.
For details see reference \cite{Z2}.} form $0 / 0$.
In the infinite-volume limit
the horizon tensor is expected \cite{Z2}
to be negative-definite inside the fundamental
modular region $\Lambda$, and to vanish on its 
boundary. In the same limit,
the measure should get concentrated \cite{Z2} on a region where
the horizon tensor per unit volume
$ h_{\mu\,a\mbox{,}\,\nu\,b}(U) $
is equal to zero. It is interesting to recall that
the condition $h(U) = 0$, where $h(U)$ is the
horizon function per unit volume,
is satisfied for all transverse configurations
on a finite lattice with free boundary conditions \cite{MZ}.
However, this result is not sufficient to make
lattices of this kind free of Gribov copies \cite{C2future}.
In order to test these conjectures, we evaluate
the 12 eigenvalues of the
horizon tensor (per unit volume) and the horizon function
(also per unit volume).
We also consider the average projection (per unit volume)
of the $ B^{c}_{\mu\,a}(\bx) $ vectors on the eigenvector
$ \omega_{min}^{c}(\bx) $, i.e.\
\be
{\cal P} \equiv
\frac{1}{12\, V}\,\sum_{\mu\mbox{,}\,a}\,
\left[\,\sum_{\bx\mbox{,}\,c}\,
B^{c}_{\mu\,a}(\bx)\,\omega_{min}^{c}(\bx)\,
    \right]^{2}
\;\mbox{,}
\label{eq:P}
\ee
and the contribution ${\cal P} / \lambda_{min}$
to the horizon function $h$ from this eigenvector
[see formulae \protect\reff{eq:hfunct2}
and \protect\reff{eq:P}].
Finally, we evaluate the largest and the
smallest (both in absolute value)
non-diagonal elements of the matrix
${\widetilde h}_{\mu\,a\mbox{,}\,\nu\,b}
\equiv 12\,h_{\mu\,a\mbox{,}\,\nu\,b}$, and the average
over its non-diagonal elements
defined as\footnote{~Here
the index $i$ (respectively $j$) stands for the pair of indices
$\mu$ and $a$ (respectively $\nu$ and $b$).}
\be
\left( \frac{1}{132}\,\sum_{i\mbox{,}\, j\mbox{;}\, i \neq j}\,
      {\widetilde h}_{i j}^{2}\,\right)^{1/2}
\;\mbox{.}
\label{eq:NDEaver}
\ee


\section{Numerical Simulations}

In order to evaluate the horizon tensor, and the horizon function,
we have to invert the Faddeev-Popov matrix ${\cal M}_{U}$
[see equations \reff{eq:hfunc} and \reff{eq:htensor}].
This matrix is rather large and sparse, and has
an eigenvalue zero
corresponding to constant modes. This means
that it cannot be inverted directly.
Nevertheless, this inversion can be done by using
a standard conjugate-gradient (CG) method, provided that we work
in the subspace orthogonal to constant vectors
(see refences \cite{Gnoise,Athesis,Ghost}).
In particular, we have to impose that the source
and the initial guess for the
CG-method have zero constant mode. In our case
the source is given by one of the
(twelve) vectors $B^{c}_{\mu\, a}(\bx)$ defined in 
eq.\ \reff{eq:Bvector}.
Thus, before inverting the matrix ${\cal M}_{U}$ we have to
impose the condition
\be
\sum_{\bx}\, B^{c}_{\mu\, a}(\bx) \,=\,0
\;\mbox{.}
\ee

The evaluation of the smallest nonzero eigenvalue
of ${\cal M}_{U}^{a\, b}( \bx\mbox{,}\,\by )$ can be done using
the routine that inverts this matrix. To this end let us
consider the vector
\be
\omega^{a}_{m}(\bx) \,\equiv\,
\sum_{\by}\,\sum_{b}\,\left(\,
{\cal M}_{U}^{- m}\,\right)^{a\, b}(\bx\mbox{,}\, \by)
\, \omega^{b}(\by) 
\;\mbox{,}
\ee
where $\omega^{b}(\by)$ is a randomly chosen vector
such that $\sum_{\by}\, \omega^{b}(\by)\,=\,0$.
Then, in the limit
of large $m$, we have
\be
\omega^{a}_{m}(\bx) \,=\,
     \,c\, \lambda^{- m}_{min}
     \, \omega^{a}_{min}(\bx) \,+\, \ldots
\label{eq:omegamin}
\;\;\mbox{,}
\ee
where $c$ is an unknown constant. This implies
\be
\lambda_{min} \,=\, \frac{\omega^{a}_{m}(\bx)}{\omega^{a}_{m+1}(\bx)}
\ee
for any $a$ and $\bx$, and $\omega^{a}_{min}(\bx)$ is given
by the vector $\omega^{a}_{m}(\bx)$ normalized to $1$.
There are of course
more sophisticated algorithms to evaluate eigenvalues and eigenvectors
of a matrix. However, for this case, this simple method converges
very fast\footnote{~This tells us that the second smallest
nonzero eigenvalue of ${\cal M}_{U}^{a\, b}( \bx\mbox{,}\,\by )$
is usually much larger than $\lambda_{min}$.}
and is easy to be implemented, since it uses the
same conjugate-gradient routine used to evaluate
the horizon tensor.

We consider, for each quantity, two different averages:
\begin{itemize}
\item Average (indicated as ``am'') only on the supposed absolute
      minima.\footnote{~We refer to \cite{Gnoise,Athesis} for details
about the numerical search for a candidate for the absolute
minimum of the minimizing function.}
\item Average (indicated as ``fc'') only on the
      first gauge-fixed gauge copy generated
      for each thermalized configuration. This is the
      result that we would obtain if Gribov copies
      were not considered.
\end{itemize}
If the result obtained from these two averages are
systematically different for a given quantity, we 
can say that the existence of Gribov copies introduces
a bias (Gribov noise) on the numerical evaluation of that quantity.

The parameters used for the simulations can be found
in \cite[Table 1]{Gnoise}. However, in this
paper, configurations at $\beta = 1.6$
and with lattice volume $V = 24^4$ were not analyzed.
Our runs were started with a randomly chosen configuration.
Details about thermalization (using hybrid overrelaxation)
and gauge-fixing (using stochastic overrelaxation) can be found
in references \cite{Gnoise,Athesis,CM}. Computations were performed on
several IBM RS-6000/250--340 workstations
at New York University.


\section{Results}

In Table \ref{FPeige} we report the data for
the ratio $\lambda_{min} / \lambda_{Lap}$, the
third and the fourth derivatives of the minimizing function,
and the ratio $r$ defined in eq.\ \reff{eq:ratiodefinizione}.
For the third derivative we consider the absolute value
since this quantity is negative for about $50 \%$ of the
configurations. The fourth derivative, on the contrary, is
always positive, with very few exceptions at $\beta = 2.7$ 
and small lattice volumes.

Our results are consistent with the conjectures discussed
in Section \ref{Sec:FMRstudio}: the value of
$\lambda_{min}$ (respectively $r$) is larger (smaller), 
in average, for the absolute minima (average ``am''). On
the contrary, Gribov noise is not observable for
the third and fourth derivatives of the minimizing
function. Thus the Gribov noise for
the ratio $r$ is entirely due to the Gribov noise
of the eigenvalue $\lambda_{min}$.
It is also interesting to observe that,
for $\beta$ in the strong-coupling regime,
$\lambda_{min}$ is down by a factor approximately $40-60$
with respect to the lowest nonzero eigenvalue $\lambda_{Lap}$
of the negative of the lattice Laplacian. (At
$\beta = 2.7$ this ratio is still about $6.5$.)
This indicates that typical (thermalized and gauge-fixed)
configurations, including those that are absolute minima,
lie very close to the Gribov horizon $\partial \Omega$,
where $\lambda_{min} = 0$.
Recall that the fundamental modular region $\Lambda$ is
included in the Gribov region $\Omega$. Consequently we
conclude that a typical configuration in $\Lambda$ lies close
to its boundary $\partial \Lambda$, and
where this boundary in turn lies close to $\partial \Omega$.

As for the quantity $r$, from Table \ref{FPeige}
we observe that (where the statistics are good)
$r$ is quite small\footnote{~We have found only in very few cases
a value of $r$ larger than $8/3$, which
indicates the existence of other stationary points
near the minimum at $\tau = 0$ (see Figure
\ref{fig:ratior}).},
even though typical configurations
lie near the boundary of the Gribov region $\partial \Omega$,
as indicated by the remarkable smallness of $\lambda_{min}
= 4\,V\,{\cal E}^{''}(0)$.
At a generic point of $\partial \Omega$ we have
${\cal E}^{'''}(0) \neq 0$, so $r = \infty$.
However, at those points of
$\partial \Omega$ that are also points of $\partial \Lambda$,
we have ${\cal E}^{'''}(0) = 0$ (see references
\cite{STSF,Z1,Z2}).
The smallness of $r$ for relative (as well as
absolute) minima is consistent with the hypothesis
that also for typical configurations
that are {\em relative} minima the boundary $\partial \Lambda$
lies close by. Note that whereas the
gauge orbit is tangent to $\Gamma \,( \partial \cdot A = 0 )$
on a generic point of $\partial \Omega$, it is also tangent to
$\partial \Omega$ at the common points of $\partial \Lambda$
and $\partial \Omega$ \cite{STSF,Z1,Z2}.

In Table \ref{horizon} we report the data for
the horizon function $h$, the contribution
${\cal P} / \lambda_{min}$ to the horizon function,
and the minimizing function ${\cal E}_{U}$.
Finally, in Table \ref{eigen_hor},
we show the values for the smallest and the largest
eigenvalues of the horizon tensor 
${\widetilde h}_{a\,\mu\mbox{,}\,b\,\nu}$,
and the average over its non-diagonal elements, as
defined in eq.\ \reff{eq:NDEaver}.

From our data there is evidence of Gribov noise for $h$ and
for the eigenvalues of the horizon tensor: these
quantities, in fact, are smaller at the absolute minima.
Moreover, the horizon tensor is closer to a diagonal matrix
at the absolute minima\footnote{~See data for
the average of the non-diagonal elements in Table
\ref{eigen_hor}.}. On the contrary there is no
evidence of Gribov noise for the quantity
${\cal P}$ defined in eq.\ \reff{eq:P}.
This suggests that the Gribov noise for ${\cal P} / \lambda_{min}$,
a quantity which contributes to the value of the
horizon function, is due to the 
eigenvalue $\lambda_{min}$. Note that
${\cal E}_{U} - 1$ contributes
to the horizon function $h$ also [see formulae 
\reff{eq:Ewith1} and \reff{eq:hfunct2}], and that
${\cal E}_{U}$ is (by definition) smallest at the absolute minimum.
However, the Gribov noise for $h$ is much larger
than the Gribov noise of ${\cal E}_{U}$ and
${\cal P} / \lambda_{min}$, i.e.\ it is probably due to
Gribov noise for {\em all} the eigenvalues of the Faddeev-Popov matrix.

As for the volume dependence, it
seems that the horizon tensor becomes
closer and closer to a diagonal matrix as the lattice size
increases (at fixed $\beta$). On the contrary, the
value of the
horizon function seems almost independent of the volume,
while ${\cal P}$ and
${\cal P} / \lambda_{min}$ decrease with $V$.
It is also evident that, at large
enough volume, the horizon tensor is negative-definite.
However, for small lattice volumes we find positive eigenvalues
of the horizon tensor. We have also found, again at
small values of $V$, a few
configurations with a positive value for the horizon
function $h$, even though $h$ is always negative
in average, and its value approaches $- 1$
(the value on the vacuum) as $\beta$ increases.

\section{Conclusions}

Our data show Gribov noise for
quantities related to the Faddeev-Popov matrix, such as
the eigenvalue $\lambda_{min}$ and the horizon tensor.
This is in agreement with the result obtained in
reference \cite{Gnoise}, in which Gribov noise has been
observed for the ghost propagator.
In all cases, the effect is small but clearly detectable
for the values of $\beta$ in the strong-coupling region.
The fact that this noise is not observable at $\beta = 2.7$
seems to us to be related only to the small volumes considered here.
Of course this hypothesis should be checked numerically.
This is, at the moment, beyond the limits of our computational resources.

As for the ``localization'' in the gauge-field space
of a {\em typical} (thermalized and gauge-fixed) configuration,
the smallness of $\lambda_{min}$ and $r$
suggest that these configurations are always close to the
common boundary of the Gribov region $\Omega$ and the
fundamental modular region $\Lambda$. This is the case
for relative as well as for absolute minima of the minimizing
function. Moreover, in the limit of large lattice volume,
only the region
\be
{\widetilde \Omega} \equiv \left\{\, A \, : \, A \in \Omega \mbox{,}\,\,
H_{\mu\,a\mbox{,}\,\nu\,b}[ U ] \, < \, 0\, \right\}
\label{eq:Omegatilde}
\ee
seems to contribute to the evaluated expectation values.
Thus our results support the conjectures \cite{Z2}
that, in the limit of large lattice volume, the horizon tensor
is negative-definite inside the fundamental
modular region, and that the measure should get concentrated
on the common boundary of $\Omega$ and $\Lambda$.


\section*{Acknowledgements}

I am indebted to D.Zwanziger for suggesting
this work to me. I would also like to thank him,
G.Dell'Antonio, T.Mendes and M.Schaden for valuable
discussions and suggestions. 

I thank for the hospitality the Physics Department 
of the University of Bielefeld and the Center
for Interdisciplinary Research (ZiF), where part of 
this work was done.


\clearpage

\vspace*{0.2cm}


%
\begin{table}
%
\hspace*{-1.0cm}
\protect\small
\begin{center}
\begin{tabular}{|| l | l | l | c | c | c | c ||}
\hline
\hline
$\beta$ & $N$ & $aver$
& $\lambda_{min} / \lambda_{Lap}$
& $4\,V\,| {\cal E}_{U}^{'''} | $
& $4\,V\,{\cal E}_{U}^{''''}$ & $r$ \\
\hline
\hline
$ 0.0 $ & $ 4 $ & am & $ 0.0251 (0.0005)
 $ & $ 0.014028 (0.000615)
 $ & $ 0.08571 (0.00259)
 $ & $ 0.083 (0.006)
 $ \\ \hline 
$ 0.0 $ & $ 4 $ & fc & $ 0.0230 (0.0005)
 $ & $ 0.015851 (0.000654)
 $ & $ 0.08960 (0.00262)
 $ & $ 0.124 (0.010)
 $ \\ \hline 
$ 0.0 $ & $ 6 $ & am & $ 0.0204 (0.0006)
 $ & $ 0.005050 (0.000369)
 $ & $ 0.03441 (0.00179)
 $ & $ 0.084 (0.015)
 $ \\ \hline 
$ 0.0 $ & $ 6 $ & fc & $ 0.0167 (0.0005)
 $ & $ 0.004918 (0.000413)
 $ & $ 0.02910 (0.00159)
 $ & $ 0.107 (0.018)
 $ \\ \hline 
$ 0.0 $ & $ 8 $ & am & $ 0.0186 (0.0006)
 $ & $ 0.003719 (0.000282)
 $ & $ 0.02287 (0.00144)
 $ & $ 0.117 (0.015)
 $ \\ \hline 
$ 0.0 $ & $ 8 $ & fc & $ 0.0162 (0.0005)
 $ & $ 0.003424 (0.000236)
 $ & $ 0.01954 (0.00143)
 $ & $ 0.141 (0.020)
 $ \\ \hline 
$ 0.0 $ & $ 10 $ & am & $ 0.0169 (0.0007)
 $ & $ 0.003025 (0.000384)
 $ & $ 0.02435 (0.00281)
 $ & $ 0.140 (0.026)
 $ \\ \hline 
$ 0.0 $ & $ 10 $ & fc & $ 0.0155 (0.0007)
 $ & $ 0.002912 (0.000455)
 $ & $ 0.01798 (0.00232)
 $ & $ 0.205 (0.044)
 $ \\ \hline 
$ 0.0 $ & $ 12 $ & am & $ 0.0178 (0.0008)
 $ & $ 0.002056 (0.000267)
 $ & $ 0.01205 (0.00110)
 $ & $ 0.155 (0.050)
 $ \\ \hline 
$ 0.0 $ & $ 12 $ & fc & $ 0.0163 (0.0007)
 $ & $ 0.002466 (0.000400)
 $ & $ 0.01565 (0.00208)
 $ & $ 0.230 (0.083)
 $ \\ \hline 
$ 0.0 $ & $ 14 $ & am & $ 0.0199 (0.0017)
 $ & $ 0.002303 (0.000484)
 $ & $ 0.01685 (0.00325)
 $ & $ 0.218 (0.078)
 $ \\ \hline 
$ 0.0 $ & $ 14 $ & fc & $ 0.0139 (0.0010)
 $ & $ 0.001605 (0.000269)
 $ & $ 0.01598 (0.00349)
 $ & $ 0.108 (0.028)
 $ \\ \hline 
$ 0.0 $ & $ 16 $ & am & $ 0.0174 (0.0018)
 $ & $ 0.002259 (0.000631)
 $ & $ 0.01925 (0.00835)
 $ & $ 0.250 (0.103)
 $ \\ \hline 
$ 0.0 $ & $ 16 $ & fc & $ 0.0157 (0.0013)
 $ & $ 0.003042 (0.000873)
 $ & $ 0.01267 (0.00274)
 $ & $ 0.453 (0.240)
 $ \\ \hline \hline
$ 0.8 $ & $ 8 $ & am & $ 0.0238 (0.0006)
 $ & $ 0.004682 (0.000541)
 $ & $ 0.02340 (0.00232)
 $ & $ 0.165 (0.035)
 $ \\ \hline 
$ 0.8 $ & $ 8 $ & fc & $ 0.0187 (0.0006)
 $ & $ 0.003929 (0.000312)
 $ & $ 0.01935 (0.00127)
 $ & $ 0.173 (0.025)
 $ \\ \hline 
$ 0.8 $ & $ 12 $ & am & $ 0.0229 (0.0015)
 $ & $ 0.002674 (0.000635)
 $ & $ 0.01197 (0.00376)
 $ & $ 0.215 (0.068)
 $ \\ \hline 
$ 0.8 $ & $ 12 $ & fc & $ 0.0178 (0.0020)
 $ & $ 0.002434 (0.000478)
 $ & $ 0.01582 (0.00457)
 $ & $ 0.245 (0.083)
 $ \\ \hline 
$ 0.8 $ & $ 16 $ & am & $ 0.0257 (0.0033)
 $ & $ 0.002092 (0.000652)
 $ & $ 0.01576 (0.00802)
 $ & $ 0.140 (0.052)
 $ \\ \hline 
$ 0.8 $ & $ 16 $ & fc & $ 0.0182 (0.0030)
 $ & $ 0.005938 (0.004303)
 $ & $ 0.02002 (0.00875)
 $ & $ 3.734 (3.600)
 $ \\ \hline \hline
$ 1.6 $ & $ 8 $ & am & $ 0.0356 (0.0013)
 $ & $ 0.003807 (0.000461)
 $ & $ 0.01391 (0.00132)
 $ & $ 0.091 (0.016)
 $ \\ \hline 
$ 1.6 $ & $ 8 $ & fc & $ 0.0277 (0.0012)
 $ & $ 0.004054 (0.000513)
 $ & $ 0.01569 (0.00144)
 $ & $ 0.205 (0.061)
 $ \\ \hline 
$ 1.6 $ & $ 16 $ & am & $ 0.0289 (0.0026)
 $ & $ 0.003869 (0.001529)
 $ & $ 0.01001 (0.00341)
 $ & $ 0.764 (0.427)
 $ \\ \hline 
$ 1.6 $ & $ 16 $ & fc & $ 0.0294 (0.0022)
 $ & $ 0.003136 (0.001060)
 $ & $ 0.01143 (0.00296)
 $ & $ 0.547 (0.300)
 $ \\ \hline \hline
$ 2.7 $ & $ 8 $ & am & $ 0.1563 (0.0051)
 $ & $ 0.000568 (0.000024)
 $ & $ 0.00027 (0.00002)
 $ & $ 0.033 (0.004)
 $ \\ \hline 
$ 2.7 $ & $ 8 $ & fc & $ 0.1552 (0.0051)
 $ & $ 0.000576 (0.000026)
 $ & $ 0.00028 (0.00002)
 $ & $ 0.036 (0.004)
 $ \\ \hline 
$ 2.7 $ & $ 12 $ & am & $ 0.1476 (0.0069)
 $ & $ 0.000144 (0.000018)
 $ & $ 0.00008 (0.00003)
 $ & $ 0.030 (0.005)
 $ \\ \hline 
$ 2.7 $ & $ 12 $ & fc & $ 0.1471 (0.0069)
 $ & $ 0.000157 (0.000019)
 $ & $ 0.00008 (0.00003)
 $ & $ 0.032 (0.005)
 $ \\ \hline 
$ 2.7 $ & $ 16 $ & am & $ 0.1492 (0.0084)
 $ & $ 0.000063 (0.000008)
 $ & $ 0.00004 (0.00003)
 $ & $ 0.026 (0.004)
 $ \\ \hline 
$ 2.7 $ & $ 16 $ & fc & $ 0.1476 (0.0085)
 $ & $ 0.000070 (0.000011)
 $ & $ 0.00006 (0.00003)
 $ & $ 0.027 (0.004)
 $ \\ \hline \hline
\end{tabular}
\end{center}
\caption{~The ratio between the
smallest
nonzero eigenvalue $\lambda_{min}$ of the
Faddeev-Popov matrix and the
smallest nonzero eigenvalue $\lambda_{Lap}$ of
(minus) the lattice Laplacian [see eq.\
\protect\reff{eq:lLap}], the absolute value
of the third derivative and the fourth
derivative of the minimizing function
${\cal E}_{U}$ [see formulae \protect\reff{eq:E3}
and \protect\reff{eq:E4}], and the ratio $r$
defined in eq.\ \protect\reff{eq:ratiodefinizione}.
The derivatives of the minimizing function are
multiplied by $4\,V$, where $V$ is the lattice volume. 
Two different types of statistics are considered:
am = absolute minimum, and fc = first copy.
Error bars (in brackets) are one standard deviation.}
\label{FPeige}
\end{table}


\clearpage

\begin{table}
%
\hspace*{-1.0cm}
\protect\small
\begin{center}
\begin{tabular}{|| l | l | l | c | c | c ||}
\hline
\hline
$\beta$ & $N$ & $aver$ & $h$ & ${\cal P} / \lambda_{min}$ & ${\cal E}_{U}$ \\
\hline
\hline
$ 0.0 $ & $ 4 $ & am & $ -0.1046 (0.0038)
 $ & $ 0.0461 (0.0022)
 $ & $ 0.37810 (0.00019)
 $ \\ \hline 
$ 0.0 $ & $ 4 $ & fc & $ -0.0940 (0.0043)
 $ & $ 0.0515 (0.0023)
 $ & $ 0.37841 (0.00019)
 $ \\ \hline 
$ 0.0 $ & $ 6 $ & am & $ -0.1377 (0.0026)
 $ & $ 0.0206 (0.0015)
 $ & $ 0.37285 (0.00012)
 $ \\ \hline 
$ 0.0 $ & $ 6 $ & fc & $ -0.1286 (0.0025)
 $ & $ 0.0241 (0.0015)
 $ & $ 0.37328 (0.00012)
 $ \\ \hline 
$ 0.0 $ & $ 8 $ & am & $ -0.1388 (0.0032)
 $ & $ 0.0126 (0.0014)
 $ & $ 0.37151 (0.00006)
 $ \\ \hline 
$ 0.0 $ & $ 8 $ & fc & $ -0.1333 (0.0018)
 $ & $ 0.0117 (0.0006)
 $ & $ 0.37193 (0.00006)
 $ \\ \hline 
$ 0.0 $ & $ 10 $ & am & $ -0.1410 (0.0011)
 $ & $ 0.0071 (0.0005)
 $ & $ 0.37117 (0.00006)
 $ \\ \hline 
$ 0.0 $ & $ 10 $ & fc & $ -0.1328 (0.0013)
 $ & $ 0.0083 (0.0008)
 $ & $ 0.37151 (0.00006)
 $ \\ \hline 
$ 0.0 $ & $ 12 $ & am & $ -0.1392 (0.0008)
 $ & $ 0.0047 (0.0004)
 $ & $ 0.37116 (0.00005)
 $ \\ \hline 
$ 0.0 $ & $ 12 $ & fc & $ -0.1342 (0.0009)
 $ & $ 0.0049 (0.0004)
 $ & $ 0.37141 (0.00005)
 $ \\ \hline 
$ 0.0 $ & $ 14 $ & am & $ -0.1387 (0.0015)
 $ & $ 0.0027 (0.0005)
 $ & $ 0.37127 (0.00007)
 $ \\ \hline 
$ 0.0 $ & $ 14 $ & fc & $ -0.1317 (0.0012)
 $ & $ 0.0043 (0.0006)
 $ & $ 0.37148 (0.00006)
 $ \\ \hline 
$ 0.0 $ & $ 16 $ & am & $ -0.1359 (0.0012)
 $ & $ 0.0029 (0.0005)
 $ & $ 0.37133 (0.00009)
 $ \\ \hline 
$ 0.0 $ & $ 16 $ & fc & $ -0.1327 (0.0018)
 $ & $ 0.0032 (0.0004)
 $ & $ 0.37150 (0.00008)
 $ \\ \hline \hline
$ 0.8 $ & $ 8 $ & am & $ -0.1618 (0.0024)
 $ & $ 0.0144 (0.0009)
 $ & $ 0.32771 (0.00008)
 $ \\ \hline 
$ 0.8 $ & $ 8 $ & fc & $ -0.1434 (0.0025)
 $ & $ 0.0175 (0.0010)
 $ & $ 0.32815 (0.00008)
 $ \\ \hline 
$ 0.8 $ & $ 12 $ & am & $ -0.1559 (0.0019)
 $ & $ 0.0049 (0.0006)
 $ & $ 0.32737 (0.00013)
 $ \\ \hline 
$ 0.8 $ & $ 12 $ & fc & $ -0.1445 (0.0040)
 $ & $ 0.0101 (0.0030)
 $ & $ 0.32770 (0.00013)
 $ \\ \hline 
$ 0.8 $ & $ 16 $ & am & $ -0.1557 (0.0019)
 $ & $ 0.0022 (0.0006)
 $ & $ 0.32720 (0.00013)
 $ \\ \hline 
$ 0.8 $ & $ 16 $ & fc & $ -0.1479 (0.0012)
 $ & $ 0.0037 (0.0006)
 $ & $ 0.32740 (0.00012)
 $ \\ \hline \hline
$ 1.6 $ & $ 8 $ & am & $ -0.1994 (0.0025)
 $ & $ 0.0181 (0.0013)
 $ & $ 0.26763 (0.00015)
 $ \\ \hline 
$ 1.6 $ & $ 8 $ & fc & $ -0.1764 (0.0047)
 $ & $ 0.0291 (0.0034)
 $ & $ 0.26797 (0.00016)
 $ \\ \hline 
$ 1.6 $ & $ 16 $ & am & $ -0.1838 (0.0027)
 $ & $ 0.0057 (0.0009)
 $ & $ 0.26643 (0.00010)
 $ \\ \hline 
$ 1.6 $ & $ 16 $ & fc & $ -0.1810 (0.0021)
 $ & $ 0.0046 (0.0008)
 $ & $ 0.26670 (0.00011)
 $ \\ \hline \hline
$ 2.7 $ & $ 8 $ & am & $ -0.4961 (0.0476)
 $ & $ 0.1378 (0.0234)
 $ & $ 0.11122 (0.00040)
 $ \\ \hline 
$ 2.7 $ & $ 8 $ & fc & $ -0.4878 (0.0482)
 $ & $ 0.1436 (0.0236)
 $ & $ 0.11128 (0.00040)
 $ \\ \hline 
$ 2.7 $ & $ 12 $ & am & $ -0.5824 (0.0164)
 $ & $ 0.0789 (0.0128)
 $ & $ 0.10101 (0.00024)
 $ \\ \hline 
$ 2.7 $ & $ 12 $ & fc & $ -0.5759 (0.0172)
 $ & $ 0.0835 (0.0133)
 $ & $ 0.10104 (0.00024)
 $ \\ \hline 
$ 2.7 $ & $ 16 $ & am & $ -0.5710 (0.0214)
 $ & $ 0.0596 (0.0096)
 $ & $ 0.09821 (0.00016)
 $ \\ \hline 
$ 2.7 $ & $ 16 $ & fc & $ -0.5764 (0.0176)
 $ & $ 0.0474 (0.0047)
 $ & $ 0.09825 (0.00016)
 $ \\ \hline \hline
\end{tabular}
\end{center}
\caption{~The horizon function (per unit volume) $h \equiv H / V$ [see
      eq.\ \protect\reff{eq:hfunc}],
      the contribution ${\cal P} / \lambda_{min}$
      to the horizon function $h$ from the eigenvector
      $ \omega_{min}^{c}(\bx) $
      [see formulae \protect\reff{eq:hfunct2}
      and \protect\reff{eq:P}], and
      the minimizing function ${\cal E}_{U}$ defined
      in eq.\ \protect\reff{eq:Ewith1}.
      Two different types of statistics are considered:
      am = absolute minimum, and fc = first copy.
      Error bars (in brackets) are one standard deviation.}
\label{horizon}
\end{table}


\clearpage

\begin{table}
%
\hspace*{-1.0cm}
\protect\small
\begin{center}
\begin{tabular}{|| l | l | l | c | c | c ||}
\hline
\hline
$\beta$ & $ N $ & $ aver $ & smallest eigen. & largest eigen.
& aver. NDE \\
\hline
\hline
$ 0.0 $ & $ 4 $ & am & $ -0.3039 (0.0018)
 $ & $ 0.4544 (0.0316)
 $ & $ 0.0613 (0.0018)
 $ \\ \hline 
$ 0.0 $ & $ 4 $ & fc & $ -0.3015 (0.0019)
 $ & $ 0.5243 (0.0348)
 $ & $ 0.0665 (0.0020)
 $ \\ \hline 
$ 0.0 $ & $ 6 $ & am & $ -0.2419 (0.0014)
 $ & $ 0.1295 (0.0220)
 $ & $ 0.0303 (0.0013)
 $ \\ \hline 
$ 0.0 $ & $ 6 $ & fc & $ -0.2394 (0.0015)
 $ & $ 0.1493 (0.0192)
 $ & $ 0.0340 (0.0013)
 $ \\ \hline 
$ 0.0 $ & $ 8 $ & am & $ -0.2105 (0.0010)
 $ & $ 0.0485 (0.0320)
 $ & $ 0.0195 (0.0012)
 $ \\ \hline 
$ 0.0 $ & $ 8 $ & fc & $ -0.2052 (0.0010)
 $ & $ 0.0346 (0.0134)
 $ & $ 0.0190 (0.0005)
 $ \\ \hline 
$ 0.0 $ & $ 10 $ & am & $ -0.1907 (0.0009)
 $ & $ -0.0427 (0.0065)
 $ & $ 0.0123 (0.0004)
 $ \\ \hline 
$ 0.0 $ & $ 10 $ & fc & $ -0.1860 (0.0010)
 $ & $ -0.0124 (0.0098)
 $ & $ 0.0139 (0.0007)
 $ \\ \hline 
$ 0.0 $ & $ 12 $ & am & $ -0.1769 (0.0007)
 $ & $ -0.0694 (0.0052)
 $ & $ 0.0090 (0.0003)
 $ \\ \hline 
$ 0.0 $ & $ 12 $ & fc & $ -0.1737 (0.0008)
 $ & $ -0.0628 (0.0043)
 $ & $ 0.0093 (0.0003)
 $ \\ \hline 
$ 0.0 $ & $ 14 $ & am & $ -0.1684 (0.0011)
 $ & $ -0.0893 (0.0063)
 $ & $ 0.0067 (0.0004)
 $ \\ \hline 
$ 0.0 $ & $ 14 $ & fc & $ -0.1635 (0.0011)
 $ & $ -0.0716 (0.0062)
 $ & $ 0.0076 (0.0004)
 $ \\ \hline 
$ 0.0 $ & $ 16 $ & am & $ -0.1603 (0.0012)
 $ & $ -0.0922 (0.0051)
 $ & $ 0.0056 (0.0003)
 $ \\ \hline 
$ 0.0 $ & $ 16 $ & fc & $ -0.1569 (0.0020)
 $ & $ -0.0887 (0.0051)
 $ & $ 0.0057 (0.0003)
 $ \\ \hline \hline
$ 0.8 $ & $ 8 $ & am & $ -0.2456 (0.0013)
 $ & $ 0.0209 (0.0146)
 $ & $ 0.0226 (0.0007)
 $ \\ \hline 
$ 0.8 $ & $ 8 $ & fc & $ -0.2362 (0.0014)
 $ & $ 0.0811 (0.0162)
 $ & $ 0.0261 (0.0008)
 $ \\ \hline 
$ 0.8 $ & $ 12 $ & am & $ -0.2035 (0.0020)
 $ & $ -0.0708 (0.0075)
 $ & $ 0.0110 (0.0005)
 $ \\ \hline 
$ 0.8 $ & $ 12 $ & fc & $ -0.1976 (0.0013)
 $ & $ -0.0132 (0.0355)
 $ & $ 0.0151 (0.0027)
 $ \\ \hline 
$ 0.8 $ & $ 16 $ & am & $ -0.1847 (0.0013)
 $ & $ -0.1158 (0.0054)
 $ & $ 0.0061 (0.0004)
 $ \\ \hline 
$ 0.8 $ & $ 16 $ & fc & $ -0.1793 (0.0015)
 $ & $ -0.0927 (0.0048)
 $ & $ 0.0073 (0.0003)
 $ \\ \hline \hline
$ 1.6 $ & $ 8 $ & am & $ -0.3158 (0.0017)
 $ & $ 0.0571 (0.0153)
 $ & $ 0.0306 (0.0010)
 $ \\ \hline 
$ 1.6 $ & $ 8 $ & fc & $ -0.3076 (0.0019)
 $ & $ 0.1985 (0.0409)
 $ & $ 0.0405 (0.0030)
 $ \\ \hline 
$ 1.6 $ & $ 16 $ & am & $ -0.2285 (0.0025)
 $ & $ -0.1055 (0.0124)
 $ & $ 0.0110 (0.0006)
 $ \\ \hline 
$ 1.6 $ & $ 16 $ & fc & $ -0.2264 (0.0017)
 $ & $ -0.1107 (0.0069)
 $ & $ 0.0104 (0.0006)
 $ \\ \hline \hline
$ 2.7 $ & $ 8 $ & am & $ -0.7818 (0.0022)
 $ & $ 1.7496 (0.5495)
 $ & $ 0.1412 (0.0220)
 $ \\ \hline 
$ 2.7 $ & $ 8 $ & fc & $ -0.7812 (0.0024)
 $ & $ 1.8130 (0.5550)
 $ & $ 0.1465 (0.0222)
 $ \\ \hline 
$ 2.7 $ & $ 12 $ & am & $ -0.7577 (0.0017)
 $ & $ 0.4859 (0.1653)
 $ & $ 0.0888 (0.0126)
 $ \\ \hline 
$ 2.7 $ & $ 12 $ & fc & $ -0.7565 (0.0019)
 $ & $ 0.5242 (0.1705)
 $ & $ 0.0910 (0.0129)
 $ \\ \hline 
$ 2.7 $ & $ 16 $ & am & $ -0.7303 (0.0029)
 $ & $ 0.2794 (0.1947)
 $ & $ 0.0708 (0.0096)
 $ \\ \hline 
$ 2.7 $ & $ 16 $ & fc & $ -0.7269 (0.0038)
 $ & $ 0.0821 (0.1000)
 $ & $ 0.0599 (0.0047)
 $ \\ \hline \hline
\end{tabular}
\end{center}
\caption{~The smallest and the largest eigenvalues
     of the matrix
     ${\widetilde h}_{\mu\,a\mbox{,}\,\nu\,b}
     \equiv 12\,H_{\mu\,a\mbox{,}\,\nu\,b} / V$
     [see eq.\ \protect\reff{eq:htensor}],
     and the average of its non-diagonal elements
     [see eq.\ \protect\reff{eq:NDEaver}].
Two different types of statistics are considered:
am = absolute minimum, and fc = first copy.
     Error bars (in brackets) are one standard deviation.}
\label{eigen_hor}
\end{table}

\end{document}